\documentclass[useAMS,usenatbib]{mn2e}
\usepackage{graphicx}
\title[A beryllium rich halo dwarf]{Possible signature of hypernova
  nucleosynthesis in a beryllium rich halo dwarf\thanks{Based on
  observations  made with ESO VLT, at Paranal Observatory, under
  programmes 076.B-0133 and on data obtained from the ESO/ST-ECF
  Science Archive Facility.}} 
\author[R. Smiljanic et al.]{R. Smiljanic$^{1,2}$\thanks{E-mail:
rodolfo@astro.iag.usp.br (RS); lpasquin@eso.org (LP)}, L.
Pasquini$^{2}$\footnotemark[2], F. Primas$^{2}$,
P. A. Mazzali$^{3,4}$, \newauthor  D. Galli$^{5}$, and G. Valle$^{6}$\\
$^{1}$Universidade de S\~ao Paulo, IAG, Dpt. de Astronomia, Rua
     do Mat\~ao 1226, S\~ao Paulo-SP 05508-090, Brazil \\
$^{2}$ European Southern Observatory,Garching bei M\"unchen, Germany
     \\
$^{3}$ Max-Planck-Institut f\"ur Astrophysik, Garching bei M\"unchen,
  Germany \\
$^{4}$ INAF - Osservatorio Astronomico de Trieste, Trieste, Italy \\
$^{5}$ INAF - Osservatorio Astrofisico di Arcetri, Firenze, Italy \\
$^{6}$ Dipartamento di Fisica, Universit\'a di Pisa, largo Pontecorvo 3, Pisa
     56127, Italy \\
}
\begin{document}

\date{Accepted 2007. Received 2007; in original form 2007}

\pagerange{\pageref{firstpage}--\pageref{lastpage}} \pubyear{2007}

\maketitle

\label{firstpage}

\begin{abstract}
 As part of a large survey of halo and thick disc stars, we found one
 halo star, HD 106038, exceptionally overabundant in beryllium. In
 spite of its low metallicity, [Fe/H] = $-$1.26, the star has
 log(Be/H) = $-$10.60, which is similar to the solar meteoritic
 abundance, log(Be/H) = $-$10.58. This abundance is more than ten
 times higher the abundance of stars with similar metallicity and
 cannot be explained by models of chemical evolution of the Galaxy
 that include the standard theory of cosmic-ray spallation. No other
 halo star exhibiting such a beryllium overabundance is known. In addition, 
overabundances of Li, Si, Ni, Y, and Ba are also observed. We suggest that 
all these chemical peculiarities, but the Ba abundance, can be simultaneously 
explained if the star was formed in the vicinity of a hypernova.

\end{abstract}


\section{Introduction}

 The single stable isotope of beryllium, $^{9}$Be, is a pure product
 of cosmic-ray spallation of heavy (mostly CNO) nuclei \citep*{RFH70}.
 Analyses of Be abundances in metal poor stars \citep{MBCP97,BDKR99}
 have found a relationship between [Fe/H]\footnote{[A/B] = log
 [N(A)/N(B)]$_{\rm \star}$ - log [N(A)/N(B)]$_{\rm\odot}$} and
 log(Be/H) with slope close to one, and between [O/H] and
 log(Be/H) with slope between 1 and 1.5, depending on the oxygen
 indicator used. Independently of the behaviour of the [O/Fe] ratio at
 lower metallicities, these results suggest a primary production of Be
 in the early Galaxy \citep{Ki01}.

 As a primary element, and assuming cosmic-rays to be globally
 transported across the early Galaxy, Be may show a smaller scatter
 than the products of stellar nucleosynthesis \citep{SY01} at a given
 time, suggesting its potential use as a cosmochronometer.

 So far, the linear relations appear to be very tight, showing a 
 surprisingly low scatter comparable to the measurement errors. This
 picture, however, might change with the increase of the samples
 analysed, as hinted by the results of \citet{BN06}. Nevertheless,
 turn-off stars of the globular clusters NGC 6397 and NGC 6752 were
 found \citep{Pas04,Pas07} to have the same Be abundance of field
 stars of the same metallicity. This strongly support the
 production of Be to be a global process. Ages derived from these
 abundances, in a comparison with a model of the evolution of Be with
 time \citep{Va02}, show an excellent agreement with ages derived from
 theoretical isochrones, supporting the use of Be as a
 cosmochronometer. Moreover, \citet{Pas05} showed that Be abundances
 could be used to study the differences in the time scales of star
 formation in the halo and the thick disc of the Galaxy. 

 In this letter, we report the discovery of an extremely Be enriched
 halo star, HD 106038, with an abundance 1.2 dex higher than stars of
 similar metallicity. This unique star deviates considerably from the
 observed relations of Be with Fe and O. It was identified during the
 analysis of a large sample containing near to one hundred halo and
 thick disc stars (Smiljanic et al. 2008, in preparation).

 Neither the standard scenario for Be production, involving spallation
 of cosmic-rays on nuclei of the interstellar medium \citep{Va02}, nor
 the superbubbles (SBs) scenario \citep{Par00} seem to be able to
  produce such Be enriched objects. The SBs model predict a scatter
 in the Be abundance \citep{PD00} that may explain the stars found by
 \citet{BDKR99} and \citet{BN06} which have similar atmospheric
 parameters but Be abundances differing by $\sim$ 0.5 dex. The very
 high Be abundance in HD 106038, however, would require an extremely
 poor mixing of the SNe ejecta with the ISM which seems to be
 difficult to justify \citep{Par00}.

\begin{figure}
\begin{centering}
\includegraphics[width=6.5cm]{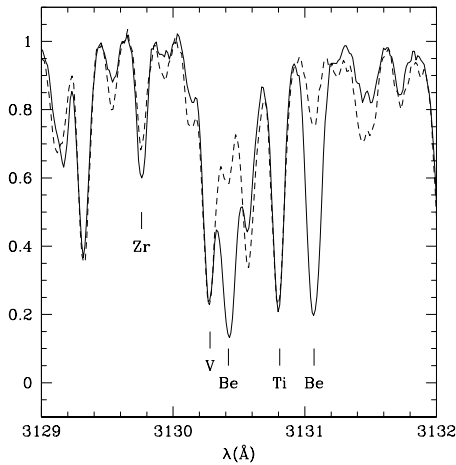}
\caption{Comparison between the spectra of HD 106038 (solid line) and
  of HIP 7459 (dashed line), a star with close atmospheric parameters
  and similar metallicity, in the Be region. The dominating element of
  the nearby blended features are also indicated. The V and Ti
  features have the same strength in the two stars while some
  difference in the Zr line is noted.}
\label{fig:be}
\end{centering}
\end{figure}

 To the best of our knowledge, there is only one other case of
 extremely Be enhanced star in the literature.\footnote{We exclude 
 from the discussion the chemically peculiar A or F stars with
 enhanced Be lines. The peculiar abundances of these stars are thought
 to be caused by  effects of diffusion. As shown by
 \citet*{Ric02}, these effects do not result in overabundances in
 stars with similar temperature and metallicity as HD 106038.} The
 star J37 of the open cluster NGC 6633 was found by \citet{As05} to 
 have log (Be/H) = $-$9.0 $\pm$ 0.5. The chemical peculiarities of
 star J37 might be best explained by the accretion of rocky material
 similar to chondritic meteorites \citep{As05}. As we shall see below,
 the accretion of such a material is unlikely for our population II
 star.

\section{Data and Analysis}

 The science raw data and calibration files of HD 106038 were
 retrieved from the ESO science archive facility. The spectra were
 originally obtained in 2000 April 12 with the UVES spectrograph 
 \citep{De00} of the ESO VLT at Cerro Paranal, Chile. The data of
 HIP 7459 (CD$-$61 282), a halo star used as comparison in this work,
 were obtained in 2005 September 22 with the same instrument. The
 spectra have R $\sim$ 45000 and a final S/N $\sim$ 70 in the Be
 region.

 For both stars, we adopt the atmospheric parameters derived by
 \citet{NS97}. The parameters were calculated with the standard
 spectroscopic approach using Fe I and Fe II lines (Table
 \ref{tab:par}).  We refer the reader to the original work for more
 details.

\begin{figure}
\begin{centering}
\includegraphics[width=6.5cm]{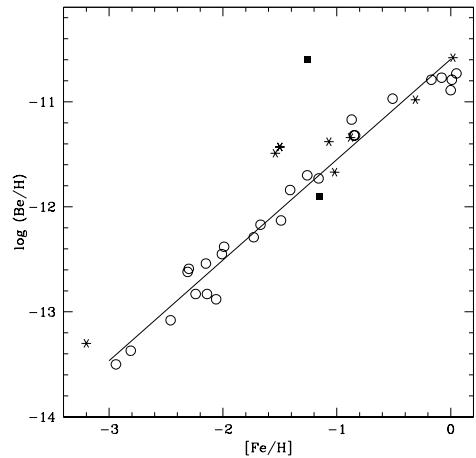}
\caption{The abundance of Be as a function of [Fe/H]. The stars HD
  106038 and HIP 7459 (filled squares) are compared to the linear
  relation defined by the stars from \citet{BDKR99} (open
  circles). The starred symbols are the stars from \citet{BDKR99} and
  \citet{BN06}. Two of them deviate from the linear relation by $\sim$
  0.50 dex.}
\label{fig:boe}
\end{centering}
\end{figure}

 Abundances (Table \ref{tab:par}) were derived through the synthesis
 of the spectrum around the Be II resonance lines at 3131.065 \AA\@
 and 3130.420 \AA. The codes for calculating synthetic spectra are
 described in \citet{Co05}. Grids of model atmospheres without
 overshooting calculated by the ATLAS9 program \citep{CK03} were
 adopted. The list of atomic lines is that compiled by \citet{Pr97}
 and the molecular line list is described in \citet{Co05}. A solar Be
 abundance was derived using the UVES solar spectrum. We estimate the
 total uncertainty from atmospheric parameters, continuum placement,
 and synthetic fit affecting the Be abundance to be $\sigma$ = $\pm$
 0.13 dex. Zero point errors might also be present due, for example,
 to the adopted model atmosphere. However, we are conducting a
 differential analysis and these should cancel out in a comparison
 between similar stars.

\begin{figure}
\begin{centering}
\includegraphics[width=6.5cm]{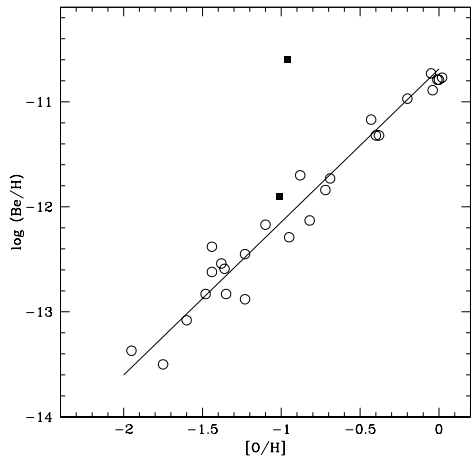}
\caption{The abundance of Be as a function of [O/H]. Symbols are as in
  Fig. \ref{fig:boe}. The stars of \citet{BN06} are not shown since O
  abundances were not derived by these authors.}  
\label{fig:boe2}
\end{centering}
\end{figure}

 A comparison between the spectra of the two stars is shown in
 Figure \ref{fig:be}, in which the extreme enhancement of the Be lines
 of HD 106038 when compared to the normal HIP 7459 is clear,
 confirming its very high Be overabundance.

 We show in Figures \ref{fig:boe} and \ref{fig:boe2} these two
 stars in the [Fe/H] vs. log (Be/H) and [O/H] vs. log (Be/H) diagrams,
 respectivelly, together with the stars from \citet{BDKR99}  and
 \citet{BN06}. Of particular interest are the two stars from
 \cite{BN06} that deviate from the linear trend. The anomalous
 position of HD 106038 clearly stands out. 

\begin{table}
\centering
\caption{The adopted atmospheric parameters and beryllium abundances
  derived using synthetic spectra for HD 106038, HIP 7459, a
  comparison star, and for the Sun.}\label{tab:par}  
\begin{tabular}{@{}cccccc@{}}
\hline
star & T$_{\rm eff}$ & log g & $\xi$ & [Fe/H] & log(Be/H) \\
 & K &  & km s$^{-1}$ &  &  \\
\hline
Sun & 5777 & 4.44 & 1.00 & 0.00 & $-$10.9 \\
HD 106038 & 6046 & 4.46 & 1.34 & $-$1.26 & $-$10.6 \\
HIP 7459 & 5909 & 4.46 & 1.23 & $-$1.15 & $-$11.9 \\
\hline
\end{tabular}
\end{table}

\subsection{Chemical abundances in the literature}

 Information on other chemical abundances might help understanding the
 origin of the Be overabundance. An overabundance of CNO elements, for
 example, would offer a good explanation for the enhancement, since
 these elements are dominant in the production of Be by spallation
 processes.

 \citet{Asp06} determined a lithium abundance of A($^{7}$Li) = 2.48
 and claimed a detection of $^{6}$Li, $^{6}$Li/$^{7}$Li =  0.031,
 compatible with the other 8 detections out of a sample of 26
 stars. The high $^{7}$Li abundance, however, results in a high
 $^{6}$Li abundance, A($^{6}$Li) = 0.97, while the mean for the other
 detections is 0.80.

 Its $^{7}$Li is particularly remarkable since it is larger than the
 Spite plateau \citep{SS82}. The plateau as found by \citet{Asp06}
 is at 2.22, which means HD 106038 has an abundance excess of
 $\Delta A$($^{7}$Li) = 2.13. Given that lithium is also expected to be
 produced by cosmic-ray spallation it seems safe to conclude that both
 $^{7}$Li and Be overabundances have the same origin.

\begin{figure}
\begin{centering}
\includegraphics[width=6.5cm]{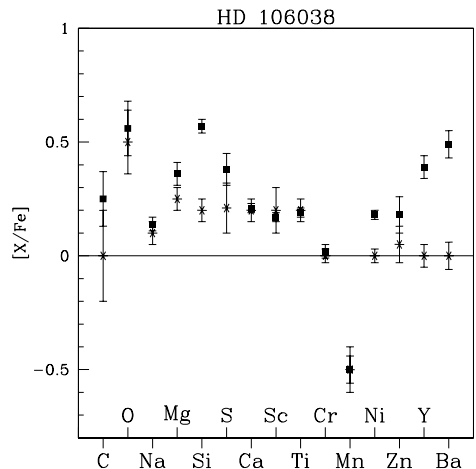}
\caption{Elemental abundances, [X/Fe], of HD 106038 determined in the
  literature. The abundances are represented by the full squares. In
  this case the error bar denotes the actual uncertainty quoted by the
  original work. The starred symbols indicate the mean abundance of
  that element for the remaining sample as found in the same work. In
  this case the error bar indicates the range of abundances for that
  given element in the original work. HD 106038 is clearly
  overabundant in Si, Ni, Y, and Ba, and show slightly larger
  abundances of C, S, Mg, and Zn when compared to the mean of the original
  samples.}
\label{fig:pat}
\end{centering}
\end{figure}

 Abundances of other elements available in the literature are shown in
 Figure \ref{fig:pat}. Abundances of Na, Ca, Mg, Si, Ti, Cr, Ni, Y,
 and Ba are from \citet{NS97}. The carbon abundance is taken from
 \citet{Fab06} and oxygen from \citet{Me06}. Abundances of Mn
 and Sc are from \citet{Nis00} and of S and Zn from \citet{Nis07}. In
 the same figure we also show the mean abundances of the samples
 analysed in these same works for comparison.

 In addition to Be and Li, the star also shows clear enhanced
 abundances of Si, Ni, and of the neutron capture elements Y and Ba. An
 enrichment in s-process elements may explain the enhanced Zr line in
 Figure \ref{fig:be}. We also note the possibly larger amounts of C,
 S, Mg, and Zn, though these remain marginally compatible with the
 mean abundances of the samples. 

\section{Discussion}

 Since the standard scenario for cosmic-ray spallation does not explain
 the enhancement of Be in HD 106038, a peculiar and/or rare event may
 be related to its formation. A combination of two or more rare
 events to produce the observed features is unlikely, we therefore
 concentrate on single events.

 To reproduce the very particular chemical pattern of HD 106038, a
 nucleosynthetic site must be able to overproduce Si and Ni without
 overproducing other $\alpha$ and iron group elements. Elements in
 normal halo stars with the same metallicity as HD106038 come mostly 
 from SNe II. It is therefore unlikely that the same SNe II may
 produce the observed enhanced [Si/Fe] and [Ni/Fe] ratios. 
 Moreover, it has about 16 times more Be than what models involving SN
 II predict for its metallicity \citep{Va02}. SNe Ia produce large
 amounts of Fe, Ni, and other iron group elements but are not expected
 to produce large amounts of O and Si \citep{Iw99}. Therefore, in case
 of a contribution from SNe Ia, ratios such as [O/Fe] and [Si/Fe]
 would fall below normal halo stars, contrary to what is
 observed. Moreover, SNe Ia are expected to produce about one
 order of magnitude less spallation products than SN II \citep{Fie02}.

 All these features, however, may possibly be found in the ejecta of a
 hypernova (HNe), which can be enriched in both intermediate mass
 elements (as S and Si) as in iron group elements
 \citep{Nak01,Pod02}. Moreover, HNe can produce large amounts of Be
 and Li by spallation \citep{Fie02,Nak04}. Therefore, we suggest that
 the material which formed this star was probably contaminated by the
 nucleosynthetic products of a HNe.

\subsection{Hypernova}\label{sec:hyp}

 Hypernovae are core-collapse SNe (usually of type Ic) with
 exceptionally large kinetic energy production, resulting in spectra
 dominated by very broad absorption line blends \citep{Maz00}. The
 energy released in the explosion can be one order of magnitude larger
 than that of normal core-collapse SNe \citep{Iw03}. Some hypernovae,
 typically the most massive and energetic events, are linked to
 Gamma-Ray bursts \citep{Iw98}.

 \citet{Fie02} and \citet{Nak04} calculated the yields of
  spallation products resulting from HNe explosions. While \citet{Nak04}
  calculate the energy distribution of the ejecta with a hydrodynamic
  code and solve the cosmic-ray transfer equation, \citet{Fie02} use
  an empirical formula for the energy distribution and do not solve
  the transfer equation but adopt an approximation to have the mass
  fraction of the ejecta that produces the spallation
  products. \citet{Nak04} claim the simplifications adopted by
  \citet{Fie02} to overestimate the yields by a factor $\sim$ 3.

 The yield of Be per HNe can be one or two orders of magnitude larger
 than the one per SNe II \citep{Fie02,Nak04}. However, as a rare
 event, they are not major contributors of Be in the
 Galaxy. \citet{Fie02} predict $^{7}$Li/$^{9}$Be $\sim$ 8.6 and Be/O
 $\sim$ $5.6 \times  10^{-7}$ (both ratios by number). The
 calculations by \citet{Nak04} predict $^{7}$Li/$^{9}$Be $\sim$ 4.2,
 also by number. Both predictions are close to what is observed in HD
 106038; the ratio between the observed excess of $^{7}$Li and
 $^{9}$Be is $^{7}$Li/$^{9}$Be = 5.6 (while the HNe is expected to
 have produced all the observed Be abundance, it is responsible only
 for the excess of Li with respect to the primordial plateau). We
 cannot estimate the contribution of the possible HNe on the observed
 oxygen abundance, thus only a lower limit can be placed, Be/O
 \textgreater $1.9 \times 10^{-7}$, given by the assumption that all
 the observed oxygen has been produced by the HNe.

 Both models, however, predict much more $^{6}$Li than
 observed, $^{7}$Li/$^{6}$Li $\sim$ 1.9 by \citet{Fie02} and
 $^{7}$Li/$^{6}$Li $\sim$ 1.2 by \cite{Nak04}, by number. The 
 observed ratio between the excess of $^{7}$Li and the $^{6}$Li
 abundance is $^{7}$Li/$^{6}$Li $\leq$ 15. We consider this ratio an
 upper limit since, given its fragility, some $^{6}$Li has probably
 been destroyed in previous evolutionary phases. The production of Be
 without a corresponding production of $^{6}$Li would be extremely
 difficult to understand.

 Nucleosynthetic calculations by \citet{Nak01} find the ejecta of HNe
 to have smaller amounts of C and O and larger amounts of Si, S, and
 Ar, when compared to normal SNe. \citet{Nak01} and \citet{No06} also
 note larger [(Zn,Co)/Fe] and smaller [(Mn,Cr)/Fe] ratios. An
 overabundance of Zn, which is not observed, can be avoided with a
 deeper mas cut\footnote{The coordinate, in mass, separating the part
 of the star that is ejected and the one that forms the remnant.},
 which would also result in a larger [Ni/Fe] \citep{No06}. These are
 in qualitatively agreement with the observations, supporting the HNe
 hypothesis.
 
 The weak s-process in massive stars seems to efficiently produce 
 only elements with a mass number\footnote{The mass number of Y is A = 
 89 and of Ba is A $\sim$ 136.} up to 90 \citep{RH00}. The Y overabundance 
 may require an enhanced flux of neutrons, which would also contribute 
 to the production of Ni. The Ba overabundance, however, is more difficult
 to understand. It is not clear whether this same mechanism would result in 
 the overproduction of Ba. Moreover, a significant amount of Ba is expected 
to be produced only by the main s-process in AGBs or by the r-process, 
 usually associated to massive stars. Since pollution by AGBs is not possible 
 (see below), the Ba overabundance is likely a product of a massive star. 
 For example, although not expected, \citet*{Maz92} found Ba to be overabundant by 
a factor of 5 in the spectrum of SN 1987A. Although Ba might pose a problem for our
 scenario, we recall that theoretical predictions for the r-process elements 
in HNe are not available. Therefore, whether this scenario is able to explain the
 Ba abundance is still an open question.

 The HNe scenario, at least qualitatively, is able to explain most
 features observed in HD 106038 within a single peculiar event. More
 work, however, is still needed to show whether the scenario still
 holds quantitatively. A detailed comparison with nucleosynthetic
 predictions of theoretical models is necessary to validate or not the
 HNe hypothesis.

\subsection{Other scenarios}\label{sec:oth}

 In this subsection we present some alternative scenarios for the origin
 of the Be enhancement in HD 106038, which were discarded for the
 reasons presented below. 

 A pollution by AGB stars, or any kind of evolved star, can at once be
 discarded. Although these may explain the s-process elements, and
 maybe Li, they do not explain the Be overabundance. On the contrary,
 the material ejected by an AGB or by a massive star, after successive
 mixing events, would be depleted in Be. 

  The SBs scenario \citep{Par00} would be able to reproduce the
  observed Li/Be and Be/O ratios only with an extreme model where
  particles are accelerated from pure SNe ejecta. The SBs evolution
  models, however, predict that most material inside of a SB comes from
  the ISM. Moreover, the remaining chemical peculiarities are not
  typical of SNe II ejecta and thus can not be explained within the
  same scenario.

 Another possibility is the engulfing of a sub-stellar object, a
 planet or planetesimals debris as in star J37 of the open cluster
 NGC 6633 \citep{As05}. This, however, can be excluded with a robust
 quantitative argument. The accreted material would be confined 
 to the surface convective layer of the star. In a metal poor star
 this layer is much shallower than in a solar metallicity star. With the
 equation given by \citet{Mu01} we estimate the surface convective
 layer of HD 106038 to have $\sim$ $4.5 \times 10^{-3}$
 M$_{\odot}$. The mass of Be in this layer is $\sim$ $7.7 \times
 10^{-13}$ M$_{\odot}$ while in a star with normal abundance of Be it
 would be $\sim$ $4.8 \times 10^{-14}$ M$_{\odot}$. Assuming the
 accreted material to have a composition similar to chondrites
 meteorites \citep{Lod03} a mass of Fe of $5.3 \times 10^{-6}$
 M$_{\odot}$ would also be accreted with the required mass of
 Be. However, the total mass of Fe in the convective layer of the star
 is $\sim$ $3.3 \times 10^{-7}$ M$_{\odot}$. In this scenario all, or
 almost all, Fe in the convective layer of the star would come from
 the accreted material. HD 106038 would then originally be a
 population III metal free star; an extremely unlikely
 possibility.  In addition, we note that the most metal poor star
 found to host planets has [Fe/H] = $-$0.68 \citep{Co07}.

 A longer exposure to EPs would be the natural explanation if HD
 106038, for some reason, was younger than halo stars of the same
 metallicity. Its Be abundance would be a result of the accumulated
 action of EPs in a cloud where star formation was, somehow,
 delayed. Its Be abundance higher than the solar photospheric
 one could be a sign of a solar or younger age, in agreement with
 the suggestion of \citet{Pas04} that Be abundances could be used as a
 cosmic clock. The position of the star in the HR diagram, although
 not favourable for a good age determination, favours an older age and
 argues against this hypothesis.

 If the star originated in or near the Galactic centre, the enhanced
 star formation and supernovae events could provide an enhanced EPs
 flux. This flux might also originate from a non-stellar source such
 as the central black hole. A bulge origin for the star, however,
 seems unlikely. In particular the abundances of Ni, Y, and Ba are not 
 compatible with the ones of bulge stars \citep{MR94}. Since it
 requires another source for the Ni and s-process overabundances, we
 also discard this hypothesis.

\section*{Acknowledgements}

This work was developed during the visit of R.S. to ESO made possible
by a CAPES fellowship (1521/06-3) and support from the ESO DGDF.

\label{lastpage}

\end{document}